\documentclass[aps,nofootinbib,notitlepage]{revtex4-1}

\usepackage{amsmath}
\usepackage{mathtools}
\usepackage{amsfonts}
\usepackage{amssymb}
\usepackage{hyperref}
\usepackage{graphicx}

\DeclareMathOperator{\tr}{tr}
\DeclareMathOperator{\rk}{rank}
\newcommand{\nn}{\nonumber}

\begin{document}

\title{Single-shot work extraction in quantum thermodynamics revisited}
\author{Shang-Yung Wang}
\email{sywang@mail.tku.edu.tw}
\affiliation{Department of Physics, Tamkang University, Tamsui District, New Taipei City 25137, Taiwan}

\begin{abstract}
We revisit the problem of work extraction from a system in contact with a heat bath to a work storage system, and the reverse problem of state formation from a thermal system state in single-shot quantum thermodynamics. A physically intuitive and mathematically simple approach using only elementary majorization theory and matrix analysis is developed, and a graphical interpretation of the maximum extractable work, minimum work cost of formation, and corresponding single-shot free energies is presented. This approach provides a bridge between two previous methods based respectively on the concept of thermomajorization and a comparison of subspace dimensions. In addition, a conceptual inconsistency with regard to general work extraction involving transitions between multiple energy levels of the work storage system is clarified and resolved. It is shown that an additional contribution to the maximum extractable work in those general cases should be interpreted \emph{not} as work extracted from the system, \emph{but} as heat transferred from the heat bath. Indeed, the additional contribution is an artifact of a work storage system (essentially a suspended ``weight'' that can be raised or lowered) that does not truly distinguish work from heat. The result calls into question the common concept that a work storage system in quantum thermodynamics is simply the quantum version of a suspended weight in classical thermodynamics.
\end{abstract}

\maketitle

\section{Introduction}

The prospect of manipulating micro- and nanoscale quantum machines~\cite{Toyabe:2010,Cheng:2012,Blickle:2012,Koskia:2014} that can harvest energy from the environment and perform work has motivated a rapid growth of the emerging field of quantum thermodynamics~\cite{Vinjanampathy:2016}. 
Thermodynamics, on the one hand, was established in the 19th century as a phenomenological theory to study macroscopic thermal machines such as steam engines. Its basic concepts have since remained unchanged, and its applicability have been extended outside the original domain. Few believe the laws of thermodynamics will ever fail. 
Quantum theory, on the other hand, was invented during the early 20th century to explain the behavior of matter at the atomic scale. The development of quantum mechanics and its comprehension form the basis of our current understanding of the physical world and the foundations of modern technology. Few doubt the universe is quantum-mechanical at the microscopic scale. 
While thermodynamics and quantum mechanics deal with phenomena at distinct, widely separated scales, the term quantum thermodynamics however is not an oxymoron. In fact, quantum thermodynamics aims to extend thermodynamics to individual quantum systems and provide a unified description of thermodynamic processes at the microscopic and macroscopic scales. 

One of the triumphs of classical thermodynamics is the explanation of how to convert a form of disordered random energy (i.e., heat) into ordered useful energy (i.e., work). However, unlike in classical thermodynamics, the notion of work in quantum thermodynamics still requires a precise definition~\cite{Frenzel:2014,Egloff:2015,Gallego:2016}. 
A common approach has been to adopt the concept of ensembles in statistical physics, and analyze the work content of a quantum system in contact with a heat bath in terms of expectation values of physical observables, such as the Hamiltonian and density matrix, on an infinite ensemble of independent and identically distributed copies of the quantum system. 
It has been shown recently in Ref.~\cite{Aberg:2013} that optimizing the expected work gain leads to intrinsic fluctuations that can be of the same order as the expected work content \emph{per se}. The extracted energy becomes unpredictable and thus intuitively more heat-like than work-like, implying the work extraction does not act as a truly ordered work-like energy source.
As an alternative, and inspired in part by single-shot information theory, the $\varepsilon$-deterministic work content of a quantum system has been introduced~\cite{Aberg:2013,Horodecki:2013}. This means that we ask what is the maximum extractable work in each single run of the extraction process, instead of the average work that can be extracted, apart from a failure probability of $\varepsilon$, not necessarily small. This scenario is referred to as the single-shot regime, as opposed to the independent and identically distributed regime.

In Ref.~\cite{Horodecki:2013}, the single-shot maximum extractable work obtained in Ref.~\cite{Aberg:2013} was generalized to the fully quantum-mechanical setting utilizing the formalism of thermal operations in the context of thermodynamic resource theory~\cite{Brandao:2013}, and the minimum work cost of formation in the single-shot regime was also derived. 
The approach used in Ref.~\cite{Horodecki:2013} relies on the concepts of min- and max-relative entropies developed in quantum information theory~\cite{Datta:2009} and a generalized majorization concept called thermomajorization. Hence, chances are that researchers, in particular those outside the field of quantum information theory, may find the reasoning and underlying physics obscured by the intricate mathematics involved. 
Recently, an alternative approach also utilizing the formalism of thermal operations but with emphasis on a more intuitive concept based on a comparison of subspace dimensions has been presented in Ref.~\cite{Gemmer:2015}. However, careful readers may find this approach unsatisfactory either, because it is not clear why the comparison of subspace dimensions has to be done in a particular way instead of another.

In this article, utilizing the formalism of thermal operations, we present a physically intuitive and mathematically simple approach to single-shot work extraction and state formation that is accessible to general readers, in particular those not familiar with quantum information theory.
The approach is physically intuitive in that it is based solely on the observation that thermal operations~\cite{Brandao:2013,Gemmer:2015} associated with state transitions in work extraction and state formation processes are implemented by global unitaries on the heat bath, system, and work storage system that strictly conserve the total energy.
Hence, the initial and final global states have the same eigenvalues (including multiplicities). The eigenvalues of the initial global state and the diagonal elements of the final global state in the global energy eigenbasis can be easily calculated, provided that the initial system state is assumed to be diagonal in its local energy eigenbasis.
 
This approach is also mathematically simple because it requires only basic knowledge of majorization theory~\cite{Marshall:2010,Gour:2015}, without invoking the more extended concept of thermomajorization, and uses only the Schur theorem in matrix analysis~\cite{Marshall:2010,Horn:2013}. 
The Schur theorem provides a comparison between the diagonal elements and the eigenvalues of a Hermitian matrix in terms of majorization, and results of the comparison are graphically represented by the Lorenz curves. The concept of $\beta$-ordering~\cite{Horodecki:2013} that is central to thermomajorization and the reason for comparing the ranks of eigenvalue and diagonal element vectors in a particular way~\cite{Gemmer:2015} arise naturally in this approach.
Therefore, our approach also builds a bridge between the two previous methods based respectively on the concept of thermomajorization~\cite{Horodecki:2013} and a comparison of subspace dimensions~\cite{Gemmer:2015}. Moreover, our approach also facilitates a detailed discussion of the necessary and sufficient conditions for both perfect and imperfect maximum work extraction, a merit not available in the previous methods of Refs.~\cite{Horodecki:2013,Gemmer:2015}.

In addition, we provide a critical examination of a conceptual inconsistency in the literature~\cite{Gemmer:2015} regarding the maximum extractable work in general work extraction involving transitions between multiple energy levels of the work storage system. 
We show that there is a violation of the second law with an additional contribution to the maximum extractable work being independent of the initial system state and increasing without bound as the density of states of the work storage system increases.
Hence, the additional contribution is more heat-like than work-like, and should count as heat transferred from the heat bath to the work storage system instead of as work extracted from the system, implying that the maximum extractable work from a system state remains unchanged in the general work extraction.
The origin of this inconsistency is a work storage system (essentially a suspended ``weight'' that can be raised or lowered~\cite{Skrzypczyk:2014}) that does not distinguish work from heat. Our result calls into question the naive notion that work in quantum thermodynamics can be reliably defined as the change in the energy of a weight.

The rest of this article is organized as follows. In Sec.~\ref{sec:preliminaries}, we describe the basic ingredients of thermodynamic resource theory and provide the mathematical background for our approach. 
The acquainted reader may skip to Sec.~\ref{sec:workextraction}, in which bounds on extractable work are derived and a graphical interpretation of the maximum extractable work and corresponding single-shot free energy is presented. 
In Sec.~\ref{sec:maxwork}, the necessary and sufficient conditions for maximum work extraction are discussed in detail, and a seeming violation of the second law concerning the final system state being nonthermal after maximum work extraction is resolved.
In Sec.~\ref{sec:workformation}, bounds on work cost of formation are derived and a graphical interpretation of the minimum work cost of formation and corresponding single-shot free energy is provided.
In Sec.~\ref{sec:generalworkextraction}, the approach is applied to general work extraction in which transitions to multiple energy eigenstates in a certain energy range of the work storage system are considered successful. In doing so, a conceptual inconsistency in the literature regarding an additional contribution to the maximum extractable work is clarified and resolved.
Finally, we summarize the findings and discuss some open questions in Sec.~\ref{sec:conclusions}.

\section{Preliminaries}\label{sec:preliminaries}

Motivated by the resource theory of quantum entanglement~\cite{Horodecki:2009}, quantum thermodynamics has been studied in recent years from a resource-theoretic perspective~\cite{Gour:2015}. In a quantum resource theory, agents are allowed to perform under a particular restriction a certain class of quantum operations for free. 
Quantum states that cannot be prepared for free are deemed resources. A resource quantum state is useful to the agents in that they can use this state to implement transformations that are outside the class of free operations. While all free states are alike, each resource state is useful in its own way. 
Hence, two central questions arise naturally in every quantum resource theory: (i) how to quantify the usefulness of different resource states, and (ii) what other resource states can one particular resource state be converted to.

A widely adopted formalism in the resource theory of quantum thermodynamics (or resource theory of athermality) is that of thermal operations~\cite{Brandao:2013}. Given a quantum system with Hamiltonian $H_S$ and a heat bath with Hamiltonian $H_B$, the free operations are completely positive, trace-preserving maps on a system state $\rho_S$ of the form
\begin{equation}
\mathcal{E}(\rho_S)\coloneqq\tr_B[V(\rho_S\otimes\tau_B)V^\dagger],
\end{equation}
or, equivalently, global unitary transformations given by 
\begin{equation}
\rho_S\otimes\tau_B\xrightarrow{~V~}\sigma_{SB},\label{eq:thermalopv}
\end{equation}
where $\tau_B$ is the thermal (i.e., Gibbs) state of the bath associated with $H_B$ at some temperature $T$, and $\sigma_{SB}$ is the final global state of the total system with $\tr_B(\sigma_{SB})=\mathcal{E}(\rho_S)$. 
In the simplest case, the local Hamiltonians at the start and the end of the transformation are assumed to be identical, and the total Hamiltonian is equal to the sum of the local Hamiltonians, i.e., $H=H_B+H_S$. 
Conservation of the total energy requires that the global unitaries $V$ commute with the total Hamiltonian $H$. It is evident that $\mathcal{E}(\tau_S)=\tau_S$, where $\tau_S$ is the thermal state of the system associated with $H_S$ at temperature $T$, i.e., a free  state. Any state other than the thermal state $\tau_S$ is a resource state. 
All of the states are classified by their free energy, and this quantity also determines the interconvertibility of states.
Over the past few years, numerous insights into quantum thermodynamics have been gained using this formalism, ranging from an improved Landauer principle~\cite{Reeb:2014}, a quantification of quantum coherence in thermodynamic processes~\cite{Lostaglio:2015}, to a family of quantum second laws~\cite{Brandao:2015}, just to name a few. 
 
In a recent development~\cite{Horodecki:2013,Skrzypczyk:2014}, a work storage system that accounts for the extraction or expenditure of work under thermal operations has been explicitly included in quantum thermodynamic resource theory. 
It acts as the quantum version of a heavy weight in classical thermodynamics~\cite{Lieb:1999}. Analogous to the lifting of a weight a certain height being identified with the work performed by a heat engine, work in the quantum regime can be thought of as the ability to excite a work storage system from one energy eigenstate to a higher one.
In the presence of a work storage system with Hamiltonian $H_W$, work extraction under thermal operations is implemented by global unitary transformations of the form [see Eq.~\eqref{eq:thermalopv}]
\begin{equation}
\rho_S\otimes\tau_B\otimes|0\rangle_W\langle 0|\xrightarrow{~V~}\sigma_{SB}\otimes|w\rangle_W\langle w|,
\end{equation}
where $|0\rangle_W$ is the ground state of the work storage system with zero energy and $|w\rangle_W$ is some excited state with energy $w>0$. Again, strict energy conservation is imposed by requiring that the global unitaries $V$ commute with the total Hamiltonian $H=H_B+H_S+H_W$. 
Moreover, to ensure that the weight cannot be used as an entropy sink, independence of the extracted work on the initial state of the weight has to be imposed~\cite{Skrzypczyk:2014}. 
This constraint is fulfilled if the global unitaries $V$ commute with translations on the weight, i.e., $[V,\Gamma_a]=0$, where $\Gamma_a$ is the translation operator on the weigh by an energy difference $a$~\cite{Skrzypczyk:2014}.
The work extracted from the system under a thermal operation is then given by the energy difference $w$ of the final and initial states of the work storage system. In practical situations, the work storage system is coupled to the system through some external control that can raise or lower the energy levels of the their Hamiltonian (for example by tuning a magnetic field) such that energy (in the form of work) is exchanged between the two systems. Decoupling of the work storage system and the system can be achieved by physically removing the latter out of the external control. 
The reader is referred to Ref.~\cite{Skrzypczyk:2014} for a detailed discussion of a physically relevant protocol of work extraction and storage.

In this article, we focus on the development of a physically intuitive and mathematically simple approach to single-shot work extraction and state formation in the thermal operation formalism, and the resolution of a conceptual inconsistency regarding the maximum extractable work in general work extraction involving transitions between multiple energy levels of the work storage system. 
We do not aim here to address the long-standing issue of complete positivity of open quantum evolution (see, e.g., Refs.~\cite{Pechukas:1994,Alicki:1995,Shaji:2005}). For the purpose of this article we will follow the basic assumptions of thermal operations used in the literature, especially those of Refs.~\cite{Horodecki:2013,Gemmer:2015}. 

In the remainder of this section, we briefly summarize the definitions of majorization and the Lorenz curve in majorization theory, and state without proof the Schur theorem in matrix analysis. The interested reader is referred to Refs.~\cite{Marshall:2010,Horn:2013,Gour:2015} for further details. In particular, a proof of the Schur theorem can be found in Refs.~\cite{Marshall:2010,Horn:2013}.

Majorization is an ordering on $d$-dimensional real vectors. For a $d$-dimensional real vector $x$, we denote by $x^\downarrow$ the $d$-dimensional vector whose components are the same as that of $x$ (including multiplicities) but rearranged in decreasing order $x^\downarrow_1\ge\ldots\ge x^\downarrow_d$.
Let $x$ and $y$ be real vectors with equal dimension $d$. We say that $x$ majorizes $y$, and write $x\succ y$, if and only if
\begin{equation}
\sum_{i=1}^n x^\downarrow_i\ge\sum_{i=1}^n y^\downarrow_i\label{eq:appmajorcond}
\end{equation}
for $n=1,\ldots,d-1$, with equality instead of inequality for $n=d$. Note that majorization does not depend on the order of the components of vectors. A closely related definition is that of weak majorization. Specifically, we say that $x$ weakly majorizes $y$ if and only if the inequality in Eq.~\eqref{eq:appmajorcond} holds for $n=1,\ldots,d$.

For vectors with nonnegative components, the majorization conditions in Eq.~\eqref{eq:appmajorcond} have a graphical representation in terms of the Lorenz curves, which were originally introduced in economics to characterize income inequality~\cite{Marshall:2010}. 
First, let us define the sum of the largest $n$ components of $x^\downarrow$ by
\begin{equation}
S_n(x^\downarrow)\coloneqq\sum_{i=1}^n x^\downarrow_i
\end{equation}
for $n=1,\ldots,d$ with $S_0(x^\downarrow)\coloneqq0$. Then, the Lorenz curve of $x$ is the piecewise continuous linear function connecting the points $(n,S_n(x^\downarrow))$ for $n=0,\ldots,d$.
Note that for the sake of convenience and presentation simplicity, the horizontal axis of the Lorenz curve in our definition is \emph{not} scaled to unity. Clearly, the Lorenz curve of $x$ is nowhere below that of $y$ if and only if $x$ majorizes $y$.
A Lorenz curve consists of line segments connected at points called elbows, where the slope of the curve changes as it passes through the points. All of the vector components contained in a single segment have the same numerical value, and hence the horizontal width of a segment is related to the multiplicity of the corresponding component.
For a Lorenz curve, the on-ramp is defined to be the first segment of the curve. 
The slope and horizontal width of the on-ramp are determined by the largest component of the corresponding vector and its multiplicity, respectively. 
Similarly, the tail of a Lorenz curve is defined to be the right-most segment where the curve is flat, and its length is related to the multiplicity of the zero component of the corresponding vector.

Historically, the first application of majorization in matrix theory is the characterization of the diagonal elements of a Hermitian matrix in terms of its eigenvalues~\cite{Marshall:2010}.
In 1932, Schur proved the following theorem: if $\lambda_1,\ldots,\lambda_d$ and $\alpha_1,\ldots,\alpha_d$ are respectively the eigenvalues and diagonal elements of a $d\times d$ Hermitian matrix (including multiplicities), then the vector of eigenvalues $\lambda$ majorizes the vector of diagonal elements $\alpha$. 
Later in 1954, Horn proved the converse of Schur's result, i.e., if $\lambda$ and $\alpha$ are two $d$-dimensional real vectors such that $\lambda$ majorizes $\alpha$, then there exists a $d\times d$ Hermitian matrix with eigenvalues $\lambda_1,\ldots,\lambda_d$ and diagonal elements $\alpha_1,\ldots,\alpha_d$. Hence, the combined theorem is usually referred to as the Schur--Horn theorem~\cite{Bengtsson:2006}. Nevertheless, we use only Schur's result in this article.
Moreover, in this article we deal with density matrices (and the projections thereof), and their vectors of eigenvalues and diagonal elements are probability vectors. A (sub)normalized probability vector is a $d$-dimensional real vector $p$ with nonnegative components $p_i\ge 0$ satisfying the (sub)normalization condition $\sum_{i=1}^d p_i=1$ ($\sum_{i=1}^d p_i<1$). 
The rank of a probability vector $p$ (density matrix $\rho$ or projector $\Pi$) is denoted by $\rk(p)$ ($\rk(\rho)$ or $\rk(\Pi)$) and defined to be the number of its nonzero components (eigenvalues). 
A probability vector or density matrix is said to have full rank if and only if its rank and dimension are equal.

\section{Maximum extractable work}\label{sec:workextraction}

We first consider the problem of single-shot work extraction from a system in contact with a heat bath to a work storage system.
Following Refs.~\cite{Horodecki:2013,Gemmer:2015}, we assume that all of the local systems have a finite-dimensional Hilbert space and a Hamiltonian with minimum energy zero, the heat bath has large but finite energy, the energy of the system is small compared with that of the heat bath, and the work storage system has no degeneracies in the energy levels (i.e., all of its energy eigenstates have different energies). 
In the simplest case under consideration, the local Hamiltonians at the start and the end of the process are assumed to be identical, and the total Hamiltonian is equal to the sum of the local Hamiltonians, i.e., $H=H_B+H_S+H_W$, where $H_B$, $H_S$, and $H_W$ are the Hamiltonians of the heat bath, system, and work storage system, respectively.
We denote the energy eigenstates of the system at energy $E_S$ by $|E_S,g_S\rangle$, where $g_S(E_S)=1,\ldots,M_S(E_S)$ is the degeneracy index with $M_S(E_S)$ being the corresponding multiplicity. Likewise, the energy eigenstates of the bath at energy $E_B$ are denoted by $|E_B,f_B\rangle$, where $f_B(E_B)=1,\ldots,M_B(E_B)$ is the degeneracy index with $M_B(E_B)$ being the corresponding multiplicity. 

Following previous studies~\cite{Horodecki:2013,Gemmer:2015}, we take the initial global state $\eta$ as a product state of the form
\begin{equation}
\eta\coloneqq\rho_S\otimes\tau_B\otimes|0\rangle_W\langle 0|.
\end{equation}
Here, $\rho_S$ is a nonthermal system state that is diagonal in the energy eigenbasis of the system, $\tau_B\coloneqq e^{-\beta H_B}/Z_B$ is the thermal state of the bath at temperature $T$, where $\beta=1/kT$ with $k$ being the Boltzmann constant and $Z_B=\tr(e^{-\beta H_B})$ is the partition function of the bath, and $|0\rangle_W$ is the lowest energy eigenstate of the work storage system (weight) with energy $E_W=0$. 
Under a thermal operation implemented by the global unitary $V$, the initial global state $\eta$ is transformed to a final global state $\eta'$, i.e., schematically we have the transformation $\eta\xrightarrow{\,V\,}\eta'$.
We consider the final global state $\eta'$ of the form
\begin{equation}
\eta'\coloneqq\sigma_{SB}\otimes\sigma_W,
\end{equation}
where $\sigma_{SB}$ is the final state of the system and bath, with the final states of the system and the bath given by $\sigma_S\coloneqq\tr_B(\sigma_{SB})$ and $\sigma_B\coloneqq\tr_S(\sigma_{SB})$, respectively. 
Accounting for single-shot work extraction with a failure probability of $\varepsilon$, where $0\le\varepsilon<1$, we take the final state of the work storage system as a mixture
\begin{equation}
\sigma_W\coloneqq\varepsilon|0\rangle_W\langle 0|+(1-\varepsilon)|w\rangle_W\langle w|,\label{eq:sigmaw}
\end{equation}
where $|w\rangle_W$ is the energy eigenstate of the work storage system with energy $E_W=w>0$. 

Our choice of the final global state $\eta'$ differs from those in the literature~\cite{Horodecki:2013,Gemmer:2015}. 
In Ref.~\cite{Horodecki:2013}, either the final system state $\sigma_S$ is in general taken to be diagonal in the system energy eigenbasis (see Supplementary Note~2 of Ref.~\cite{Horodecki:2013}), or the final state of the system and bath $\sigma_{SB}$ is in specific taken to be a product state $\sigma_S\otimes\tau_B$ (see Supplementary Note~4 of Ref.~\cite{Horodecki:2013}), which however is not assumed in our analysis. Moreover, while a mixed state for the weight formally similar to $\sigma_W$ is also considered in Ref.~\cite{Horodecki:2013}, the weight state corresponding to an \emph{unsuccessful} work extraction process is not the ground state $|0\rangle_W\langle 0|$, but some (unspecified) mixed state orthogonal to $|w\rangle_W\langle w|$ and diagonal in the energy eigenbasis of the weight (see Supplementary Note~4 of Ref.~\cite{Horodecki:2013}). This implies that a failure of work extraction could still extract an arbitrary amount of work, albeit in an amount different from $w$. 
Hence, by a failure of work extraction we mean in the strict sense of the term that there is no work being extracted from the system at all. As we will see in Sec.~\ref{sec:generalworkextraction}, this strict sense of work extraction failure is in agreement with the idea developed in Ref.~\cite{Egloff:2015} that only predictable energy transfer (with failure probability $\varepsilon$) should count as work, anything beyond that as heat. 
In Ref.~\cite{Gemmer:2015}, as what we have done here, the final state of the system and bath $\sigma_{SB}$ is not assumed to be a product state $\sigma_S\otimes\tau_B$. 
However, the final weight state $\sigma_W$ is taken to be a pure state $|w\rangle_W\langle w|$; hence, a failure probability of $\varepsilon$ for imperfect work extraction had to be put in by hand so as to exclude the tails of weight $\varepsilon$ in the initial global state $\eta$. 
As will be seen below, our choice of the final weight state $\sigma_W$ not only is physically motivated but also provides a justification of the method presented in Ref.~\cite{Gemmer:2015}.

Since the global unitary $V$ commutes with the total Hamiltonian $H$, it is strictly energy conserving, and hence we may treat the transformation $\eta\xrightarrow{\,V\,}\eta'$ for each energy shell at total energy $E$ separately and combine their contributions in the end. 
Thus, it is convenient to define the projector onto the energy shell $E$ by
\begin{equation}
\Pi^E\coloneqq\sum_{E_S,E_W}\Pi_S(E_S)\otimes\Pi_B(E-E_S-E_W)\otimes\Pi_W(E_W),
\end{equation}
where $\Pi_S(E_S)$, $\Pi_B(E_B)$, and $\Pi_W(E_W)$ are projectors of the system, bath, and weight onto the local states with corresponding energies $E_S$, $E_B$, and $E_W$, respectively. The initial and final global states in the energy shell at total energy $E$ are then given by $\eta^E=\Pi^E\eta\Pi^E$ and $\eta'^E=\Pi^E\eta'\Pi^E$, respectively.
For later convenience, we denote by $d^E$ and $P^E$ the dimension of the energy shell at total energy $E$ and the probability that the initial global state $\eta$ falls in the energy shell $E$, respectively, i.e., we have $d^E=\rk(\Pi^E)$ and $P^E=\tr(\eta^E)$ with $0\le P^E\le 1$ and $\sum_E P^E=1$.

The eigenvalues of the initial global state $\eta$ in the energy shell $E$ (i.e., the projective state $\eta^E$) are given by
\begin{equation}
r^E(E_S,g_S,f_B,E_W)=\lambda(E_S,g_S)\frac{e^{-\beta(E-E_S)}}{Z_B}\delta_{E_W,0},\label{eq:reext}
\end{equation}
where $\lambda(E_S,g_S)$ are the eigenvalues of the initial system state $\rho_S$, which is taken to be diagonal in the system energy eigenbasis, with $E_S$ being the system energy and $g_S=1,\ldots,M_S(E_S)$ the corresponding degeneracy index, and $f_B=1,\ldots,M_B(E-E_S-E_W)$ is the degeneracy index of the bath at energy $E_B=E-E_S-E_W$. 
The bath multiplicity $M_B(E_B)$ is assumed to scale exponentially with $E_B$, and, following Refs.~\cite{Horodecki:2013,Gemmer:2015}, we will henceforth use the approximation $M_B(E-E_S)=M_B(E)e^{-\beta E_S}$ for $E\gg E_S$.
We note that since the global unitary $V$ preserves probability, it follows that the eigenvalues of the final global state $\eta'$ in the energy shell $E$ (i.e., the projective state $\eta'^E$) are also given by $r^E(E_S,g_S,f_B,E_W)$, and the probability that the final global state $\eta'$ falls in the energy shell $E$ is also equal to $P^E$, i.e., $\tr(\eta'^E)=P^E$.
While the initial global state $\eta$ is diagonal in the global energy eigenbasis that defines the energy shells, because the total Hamiltonian $H$ has degeneracies the final global state $\eta'$ in general is not diagonal in this basis. 
However, since the global unitary $V$ commutes with the total Hamiltonian $H$, the final global state $\eta'$ is block-diagonal in the global energy eigenbasis with no off-diagonal elements between energy shells at different total energies. 
Consequently, the final system state $\sigma_S$ may have off-diagonal elements within the energy subspace $E_S$, i.e., $\langle E_S,g_S|\sigma_{S}|E_S,g'_S\rangle\ne 0$.
The diagonal elements of the projective state $\eta'^E$ in the global energy eigenbasis are given by
\begin{align}
s^E(E_S,g_S,f_B,E_W)&=\varepsilon\langle E_S,g_S,E,f_B,0|\sigma_{SB}|E_S,g_S,E,f_B,0\rangle\delta_{E_W,0}+(1-\varepsilon)\nn\\
&\quad\times\langle E_S,g_S,E,f_B,w|\sigma_{SB}|E_S,g_S,E,f_B,w\rangle\delta_{E_W,w}.\label{eq:se}
\end{align}
Here, $|E_S,g_S,E,f_B,w\rangle$ is a shorthand notation for $|E_S,g_S\rangle{|E-E_S-w,f_B\rangle}$. It is noted that the eigenvalues $r^E(E_S,g_S,f_B,E_W)$ and diagonal elements $s^E(E_S,g_S,f_B,E_W)$ of the state $\eta'^E$ are nonnegative and separately add up to $P^E$, i.e.,
\begin{subequations}\label{eq:subnormconds}
\begin{align}
&\sum_{E_S,g_s}\sum_{f_B,E_W}r^E(E_S,g_S,f_B,E_W)=P^E,\\
&\sum_{E_S,g_s}\sum_{f_B,E_W}s^E(E_S,g_S,f_B,E_W)=P^E,
\end{align}
\end{subequations}
and hence they can be thought of as the components of $d^E$-dimensional subnormalized probability vectors of eigenvalues $r^E$ and diagonal elements $s^E$, respectively.

The Schur theorem implies that the probability vector $r^E$ majorizes the probability vector $s^E$.
To apply the Schur theorem to $r^E$ and $s^E$, their components need to be rearranged in decreasing order. 
Let $r^{E\downarrow}$ and $s^{E\downarrow}$ denote respectively the vectors having the same components as $r^E$ and $s^E$ (including multiplicities) but rearranged such that the components are in decreasing order. Specifically, we have $r^{E\downarrow}_1\ge\cdots\ge r^{E\downarrow}_{d^E}$, where $r^{E\downarrow}_1$ is the $r^E(E_S,g_S,f_B,0)$ with the largest $\lambda(E_S,g_S)e^{\beta E_S}$, $r^{E\downarrow}_2$ is that with the second largest $\lambda(E_S,g_S)e^{\beta E_S}$, and so on, until that with the smallest nonzero $\lambda(E_S,g_S)e^{\beta E_S}$ is reached, and then the remaining are all zeros. Note that such an ordering of the eigenvalues $r^E$ (and hence $\lambda$) is referred to as $\beta$-ordering in the context of thermomajorization~\cite{Horodecki:2013}.
Next, to find $s^{E\downarrow}$ we need to determine the magnitude of the diagonal elements of $\sigma_{SB}$ on the right-hand side of  Eq.~\eqref{eq:se}. 
To this end, we assume that the final bath state $\sigma_B=\tr_{S}(\sigma_{SB})$ is element-wise closed to the thermal state $\tau_B$, a plausible assumption for a large but finite bath and a small system. 
Indeed, in the context of quantum typicality~\cite{Goldstein:2006,Reimann:2007,Popescu:2006}, it has been shown~\cite{Gemmer:2015} that a typical final system state $\sigma_S=\tr_{B}(\sigma_{SB})$ is element-wise close to the thermal state $\tau_S$ if the bath is large enough but finite (see Sec.~\ref{sec:maxwork}).
Hence, the diagonal elements ${\langle E_S,g_S,E,f_B,E_W|}\sigma_{SB}{|E_S,g_S,E,f_B,E_W\rangle}$ in Eq.~\eqref{eq:se} are independent of the bath degeneracy index $f_B=1,\ldots,M_B(E-E_S-E_W)$.
Then, using the subnormalization condition for $s^E$ in Eq.~\eqref{eq:subnormconds} and the fact that
\begin{subequations}\label{eq:sumoversew}
\begin{align}
&\sum_{E_S,g_s,f_B}s^E(E_S,g_S,f_B,0)=\varepsilon P^E,\\
&\sum_{E_S,g_s,f_B}s^E(E_S,g_S,f_B,w)=(1-\varepsilon)P^E,
\end{align}
\end{subequations}
we obtain
\begin{align}
&\sum_{E_S,g_s}e^{-\beta E_S}\bigl[\langle E_S,g_S,E,f_B,0|\sigma_{SB}|E_S,g_S,E,f_B,0\rangle\nn\\
&\qquad\qquad\qquad 
-e^{-\beta w}\langle E_S,g_S,E,f_B,w|\sigma_{SB}|E_S,g_S,E,f_B,w\rangle\bigr]=0,\label{eq:sumesw}
\end{align}
where use has been made of $M_B(E-E_S-w)=M_B(E)e^{-\beta(E_S+w)}$ for $E\gg E_S+w$. Since Eq.~\eqref{eq:sumesw} is a very general relation that is independent of $w$ and $\sigma_{SB}$, we conclude that each term in the summation vanishes identically. This implies that for all values of $E_S$, $g_S$, and $f_B$, we have
\begin{equation}
\langle E_S,g_S,E,f_B,0|\sigma_{SB}|E_S,g_S,E,f_B,0\rangle=e^{-\beta w}\langle E_S,g_S,E,f_B,w|\sigma_{SB}|E_S,g_S,E,f_B,w\rangle.
\end{equation}
Thus, for $w>0$, it follows that either $s^E(E_S,g_S,f_B,w)$ and $s^E(E_S,g_S,f_B,0)$ both vanish, or 
\begin{equation}
s^E(E_S,g_S,f_B,w)>s^E(E_S,g_S,f_B,0)>0\label{eq:sewgesezero}
\end{equation}
for $0\le\varepsilon<1/(1+e^{-\beta w})$.
Assuming the amount of work $w$ that can be extracted is on the order of $kT$, we have $\beta w\approx 1$ and hence $\varepsilon\lesssim 0.73$. This condition is satisfied for physically interesting cases in which the failure probability $\varepsilon$ is \emph{not} expected to be large. 
Therefore, we have $s^{E\downarrow}_1\ge\cdots\ge s^{E\downarrow}_{d^E}$, where $s^{E\downarrow}_1$ is the largest of $s^E(E_S,g_S,f_B,w)$, $s^{E\downarrow}_2$ is the second largest of $s^E(E_S,g_S,f_B,w)$, and so on, until all of the nonzero $s^E(E_S,g_S,f_B,w)$ have been exhausted, then repeating the process for $s^E(E_S,g_S,f_B,0)$, and finally the remaining are all zeros.
In other words, the components of $s^{E\downarrow}$ can be divided into three groups: (i) nonzero $s^E(E_S,g_S,f_B,w)$ in decreasing order, (ii) nonzero $s^E(E_S,g_S,f_B,0)$ in decreasing order, and (iii) zeros. 
The first group contains the nonzero probabilities of the global energy eigenstates in the energy shell $E$ after a \emph{successful} work extraction process (with probability $1-\varepsilon$); the second group contains those after an \emph{unsuccessful} work extraction process (with probability $\varepsilon$); the third group contains the global energy eigenstates in the energy shell $E$ that will never be populated after a work extraction process, be it successful or otherwise.
We will denote by $s^E_w$ the \emph{full-rank} subnormalized probability vector whose components are nonzero $s^E(E_S,g_S,f_B,w)$ in decreasing order of the first group, i.e., the nonzero probabilities of the global energy eigenstates in the energy shell $E$ after a successful work extraction process.
Clearly, the components of $s^E_w$ add up to $(1-\varepsilon)P^E$ because of the subnormalization condition for $s^E(E_S,g_S,f_B,w)$ in Eq.~\eqref{eq:sumoversew}.

\begin{figure}[t]
\begin{center}
\includegraphics[width=4.25in]{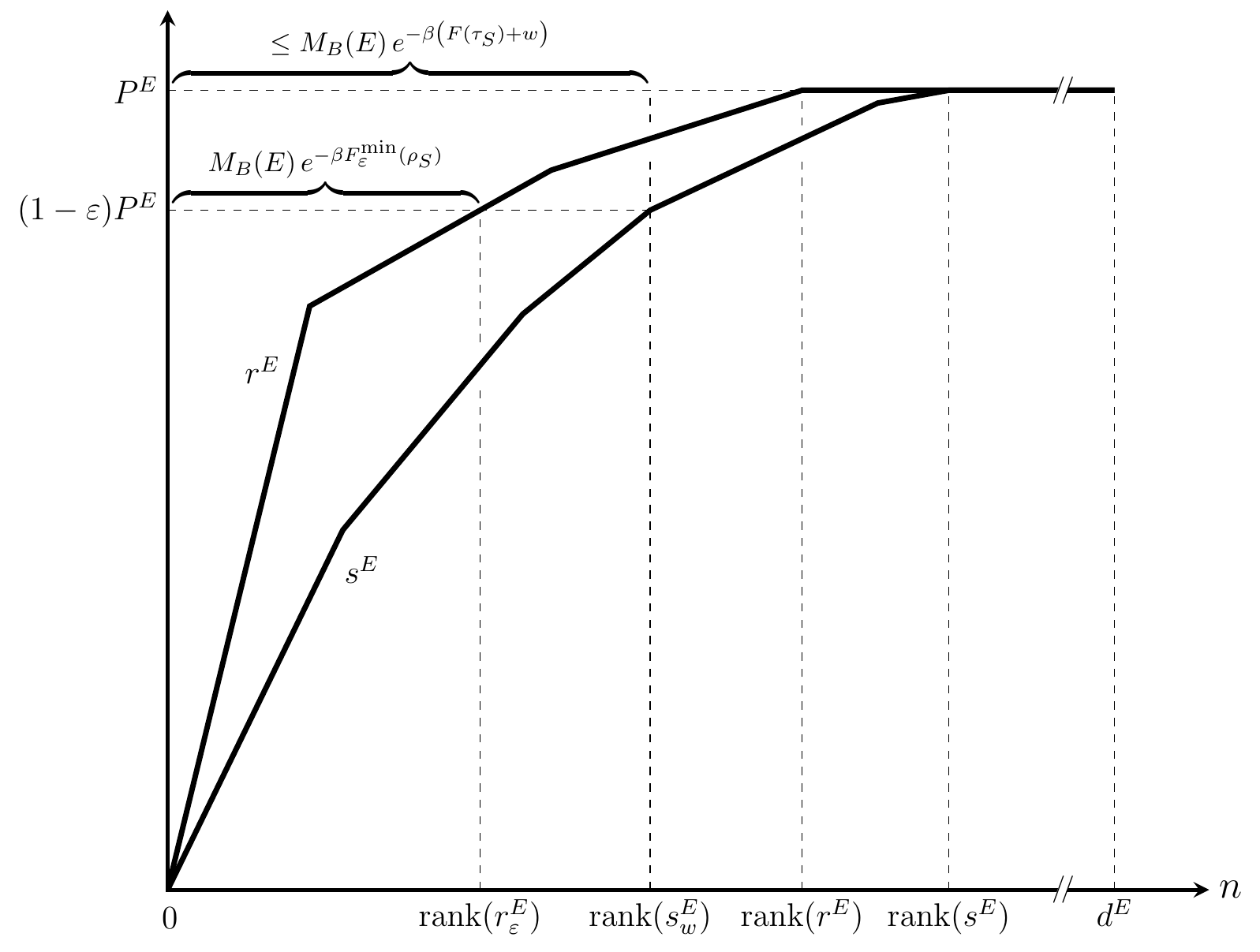}
\end{center}
\caption{Schematic Lorenz curves of $r^E$ and $s^E$ for imperfect work extraction, and graphical interpretation of the extractable work $w$ and single-shot free energy $F^\mathrm{min}_\varepsilon(\rho_S)$. The horizontal axis is broken at $n\gtrsim\rk(s^E)$ for better illustration.}
\label{fig:fig1}
\end{figure}

The fact that $r^E$ majorizes $s^E$ implies that $r^{E\downarrow}$ and $s^{E\downarrow}$ satisfy the following majorization inequalities
\begin{equation}
\sum_{i=1}^n r^{E\downarrow}_i\ge\sum_{i=1}^n s^{E\downarrow}_i\label{eq:majorcond}
\end{equation}
for $n=1,\ldots,d^E-1$, with equality for $n=d^E$ because of the subnormalization conditions in Eq.~\eqref{eq:subnormconds}.
A schematic of the Lorenz curves of $r^E$ and $s^E$ is shown in Fig.~\ref{fig:fig1}. 
Clearly, $r^E$ majorizes $s^E$ if and only if  the Lorenz curve of $r^E$ is nowhere below that  of $s^E$. 
Note that since $d^E\sim O(e^{\beta E})\gg 1$, a quasicontinuous approximation in which $n/d^E$ is effectively treated as a quasicontinuous variable can be safely employed for a large but finite bath and a small system, and will be used henceforth for presentational convenience.  

We are now in a position to identify the nonzero \emph{initial} probabilities of the global energy eigenstates in the energy shell $E$ that will be transferred entirely to $s^E_w$ with a probability of $1-\varepsilon$ in a work extraction process. 
As can be seen from Fig.~\ref{fig:fig1}, these nonzero probabilities correspond to the \emph{largest} nonzero components of $r^{E\downarrow}$ that add up to $(1-\varepsilon)P^E$. Specifically, for a given $\varepsilon\ge 0$, there exists a positive integer $n_\varepsilon\le\rk(r^E)$ such that\footnote{Here, as noted above, we have taken the quasicontinuous approximation. Otherwise, $n_\varepsilon\le\rk(r^E)$ is the \emph{smallest} positive integer such that $\sum_{i=1}^{n_\varepsilon}r^{E\downarrow}_i\ge(1-\varepsilon)P^E$, and the arguments that lead to the maximum extractable work given by Eq.~\eqref{eq:wlewmax} remain valid.}
\begin{equation}
\sum_{i=1}^{n_\varepsilon}r^{E\downarrow}_i=(1-\varepsilon)P^E,\label{eq:sumredownarrow}
\end{equation}
where $\rk(r^E)$ is the number of nonzero components of $r^E$.
According to Eq.~\eqref{eq:sumredownarrow}, we will denote by $r^E_\varepsilon$ the \emph{full-rank} subnormalized probability vector whose components are the largest $n_\varepsilon$ nonzero components of $r^{E\downarrow}$ that add up to $(1-\varepsilon)P^E$, or equivalently, the nonzero probabilities of the global energy eigenstates in the energy shell $E$ that will be transferred entirely to $s^E_w$ with a probability of $1-\varepsilon$ in a work extraction process. Clearly, we have $n_\varepsilon=\rk(r^E_\varepsilon)$ as can be seen from Fig.~\ref{fig:fig1}. 

To find the positive integer $n_\varepsilon$, or equivalently $\rk(r^E_\varepsilon)$, we first construct  a generalized ``indicator function'' $h_\varepsilon(E_S,g_S)$ that takes values in the interval $[0,1]$. A value of one (zero) for $h_\varepsilon(E_S,g_S)$ indicates that $r^E(E_S,g_S,f_B,0)$ is (not) a component of $r^E_\varepsilon$ for \emph{all} of the bath degeneracy index $f_B=1,\ldots,M_B(E-E_S)$, while a fractional value of $h_\varepsilon(E_S,g_S)$ indicates that $r^E(E_S,g_S,f_B,0)$ is a component of $r^E_\varepsilon$ for a fraction of the bath degeneracy index $f_B=1,\ldots,M_B(E-E_S)$.
Then, in terms of $h_\varepsilon(E_S,g_S)$, $\rk(r^E_\varepsilon)$ can be expressed as
\begin{align}
\rk(r^E_\varepsilon)&=\sum_{E_S,g_S}h_\varepsilon(E_S,g_S)M_B(E-E_S)\nn\\
&=M_B(E)\sum_{E_S,g_S}h_\varepsilon(E_S,g_S)e^{-\beta E_S}.\label{eq:rankreeps}
\end{align}
Moreover, as can be seen from Fig.~\ref{fig:fig1}, the fact that $r^E$ majorizes $s^E$ implies
\begin{equation}
\rk(r^E_\varepsilon)\le\rk(s^E_w),\label{eq:rkrelesew}
\end{equation}
where $\rk(s^E_w)$ is upper bounded by the number of the diagonal elements $s^E(E_S,g_S,f_B,w)$ (or the dimension of the projective subspace $\Pi_W(w)$), i.e.,
\begin{align}
\rk(s^E_w)&\le\dim(\Pi_W(w))\nn\\
&=\sum_{E_S,g_S}M_B(E-E_S-w)\nn\\
&=M_B(E)e^{-\beta w}Z_S.\label{eq:ranksee}
\end{align}
Here, $Z_S=\sum_{E_S}M_S(E_S)e^{-\beta E_S}$ is the partition function of the system at temperature $T$.
Upon combining Eqs.~\eqref{eq:rankreeps}--\eqref{eq:ranksee}, we obtain
\begin{equation}
w\le w^\mathrm{max}_\varepsilon\coloneqq F^\mathrm{min}_\varepsilon(\rho_S)-F(\tau_S),\label{eq:wlewmax}
\end{equation}
where $F^\mathrm{min}_\varepsilon(\rho_S)$ is the single-shot free energy of the (nonthermal) system state $\rho_S$ associated with a failure probability of $\varepsilon$, and defined by\footnote{In resource-theoretic terms, $F^\mathrm{min}_\varepsilon(\rho_S)$ is called the $\varepsilon$-smooth min-free energy of the state $\rho_S$~\cite{Horodecki:2013}, a quantity related to the smooth min-relative entropy of $\rho_S$ to $\tau_S$~\cite{Gour:2015,Datta:2009}, and hence the notation.}
\begin{equation}
F^\mathrm{min}_\varepsilon(\rho_S)\coloneqq-kT\log\sum_{E_S,g_S}h_\varepsilon(E_S,g_S)e^{-\beta E_S},
\end{equation}
$\tau_S\coloneqq e^{-\beta H_S}/Z_S$ is the thermal system state at temperature $T$, and $F(\tau_S)\coloneqq-kT\log Z_S$ is the standard free energy of the system. Note that $w^\mathrm{max}_\varepsilon\ge 0$ because $F^\mathrm{min}_\varepsilon(\rho_S)\ge F(\tau_S)$.
One of the merits of our approach is that, as shown in Fig.~\ref{fig:fig1}, it allows for a graphical interpretation of the extractable work $w$ and single-shot free energy $F^\mathrm{min}_\varepsilon(\rho_S)$ in terms of the Lorenz curves of $r^E$ and $s^E$. 
An inspection of Fig.~\ref{fig:fig1} shows that $w^\mathrm{max}_\varepsilon$ is indeed the least upper bound of $w$.
Moreover, $w^\mathrm{max}_\varepsilon$ is independent of the energy shell $E$ because the trivial dependence of $M_B(E)$ on $E$ cancels out; hence, the maximum extractable work previously derived in Refs.~\cite{Horodecki:2013,Gemmer:2015} is recovered. The conditions for maximum work extraction will be discussed in Sec.~\ref{sec:maxwork}.

For the special case of perfect work extraction, i.e., $\varepsilon=0$, we have $\rk(r^E_\varepsilon)=\rk(r^E)$. From Eqs.~\eqref{eq:rkrelesew} and \eqref{eq:ranksee}, we obtain 
\begin{equation}
\rk(r^E)\le\rk(s^E_w)\le M_B(E)e^{-\beta w}Z_S,\label{eq:rkrerkse}
\end{equation}
where  
\begin{equation}
\rk(r^E)=M_B(E)\sum_{E_S,g_S}h_0(E_S,g_S)e^{-\beta E_S}.\label{eq:rkrembe}
\end{equation} 
If $h_0(E_S,g_S)=1$ for all values of $E_S$ and $g_S$, then $r^E$ has full rank in the projective subspace $\Pi_W(0)$ with $\rk(r^E)= M_B(E)Z_S$ [see Eq.~\eqref{eq:rkrembe}]. This, together with Eq.~\eqref{eq:rkrerkse}, implies that $w\le 0$, and hence no work can be extracted. The same result can also be obtained by noting that if $h_0(E_S,g_S)=1$ for all values of $E_S$ and $g_S$, then $F^\mathrm{min}_0(\rho_S)=F(\tau_S)$ and hence $w\le 0$. 
Thus, with certainty (i.e., $\varepsilon=0$) work can only be extracted if the vector of eigenvalues $r^E$ of the initial projective state $\eta^E$ in any energy shell $E$ does not have full rank in the projective subspace $\Pi_W(0)$, or equivalently, the initial system state $\rho_S$ does not have full rank. 

\section{Conditions for maximum work extraction}\label{sec:maxwork}

Having established the bounds on single-shot work extraction with a failure probability of $\varepsilon$, we now study the conditions for maximum work extraction. 
Indeed, compared with previous methods, another merit of our approach is that it allows for an unambiguous identification of the necessary and sufficient conditions for both perfect and imperfect maximum work extraction.

The equality in Eq.~\eqref{eq:rkrelesew} holds if and only if $r^E_\varepsilon$ and $s^E_w$ have equal rank (or dimension) $n_\varepsilon$ and equal subnormalization $(1-\varepsilon)P^E$, or equivalently there exists a positive integer $n_\varepsilon$ such that\footnote{Here, again, the quasicontinuous approximation has been taken. Otherwise, $r^E_\varepsilon$ and $s^E_w$ have equal rank (or dimension) $n_\varepsilon$, but may not have the same subnormalization, i.e., $n_\varepsilon$ is the positive integer such that $\sum_{i=1}^{n_\varepsilon}r^{E\downarrow}_i\ge\sum_{i=1}^{n_\varepsilon}s^{E\downarrow}_i=(1-\varepsilon)P^E$. Accordingly, the conditions for maximum work extraction remain valid, except that $r^E_\varepsilon$ \emph{weakly} majorizes $s^E_w$.}
\begin{equation}
\sum_{i=1}^{n_\varepsilon}r^{E\downarrow}_i=\sum_{i=1}^{n_\varepsilon}s^{E\downarrow}_i=(1-\varepsilon)P^E.\label{eq:sumremaxext}
\end{equation}
Clearly, $n_\varepsilon$ is equal to the rank of $s^E_w$, i.e., $n_\varepsilon=\rk(s^E_w)$.
Together with Eq.~\eqref{eq:majorcond}, Eq.~\eqref{eq:sumremaxext} implies that $r^E_\varepsilon$ majorizes $s^E_w$. Moreover, the equality in Eq.~\eqref{eq:ranksee} holds if and only if all of the diagonal elements $s^E(E_S,g_S,f_B,w)$ are nonzero, i.e., $s^E_w$ has full rank in the projective subspace $\Pi_W(w)$. From Eq.~\eqref{eq:sewgesezero}, this in turn implies that all of the diagonal elements $s^E(E_S,g_S,f_B,0)$ are nonzero as well, and hence $s^E$ has full rank in the projective subspace $\Pi_W(0)+\Pi_W(w)$. 
Therefore, for a given $\varepsilon>0$ and some fixed initial system state $\rho_S$, the \emph{imperfect} maximum extractable work $w^\mathrm{max}_\varepsilon$ can be achieved if and only if, in any energy shell $E$, the vector of diagonal elements $s^E_w$ of the final projective state $\eta'^E$ has full rank in the projective subspace $\Pi_W(w^\mathrm{max}_\varepsilon)$, and is majorized by the vector of the largest $n_\varepsilon=\rk(s^E_w)$ nonzero eigenvalues $r^E_\varepsilon$ of the initial projective state $\eta^E$. Moreover, after a maximum work extraction the decohered (or dephased) final system state $\mathcal{D}(\sigma_S)$ has full rank, where $\sigma_S=\tr_B(\sigma_{SB})$ and $\mathcal{D}$ is the operation that removes all of the off-diagonal elements (i.e., quantum coherence) in the system energy eigenbasis.

The case of perfect maximum work extraction requires special care. From Eq.~\eqref{eq:rkrerkse}, maximum work extraction entails that $\rk(r^E)=\rk(s^E)$ and $s^E$ has full rank in the projective subspace $\Pi_W(w)$. 
Therefore, for some fixed initial system state $\rho_S$, the \emph{perfect} maximum extractable work $w^\mathrm{max}_0$ can be achieved if and only if, in any energy shell $E$, the vector of diagonal elements $s^E$ of the final projective state $\eta'^E$ has full rank in the projective subspace $\Pi_W(w^\mathrm{max}_0)$, and the vector of eigenvalues $r^E$ of the initial projective state $\eta^E$ has the same rank as $s^E$, but does \emph{not} have full rank in the projective subspace $\Pi_W(0)$. 
We note that since the dimension of the projective subspace $\Pi_W(w)$ in the energy shell $E$ scales as $e^{-\beta\omega}$ [see Eq.~\eqref{eq:ranksee}], the condition that $r^E$ does not have full rank in the projective subspace $\Pi_W(0)$ can always be satisfied. Moreover, like in imperfect maximum work extraction, the decohered final system state $\mathcal{D}(\sigma_S)$ after a perfect maximum work extraction also has full rank.

\begin{figure}[t]
\begin{center}
\includegraphics[width=4.25in]{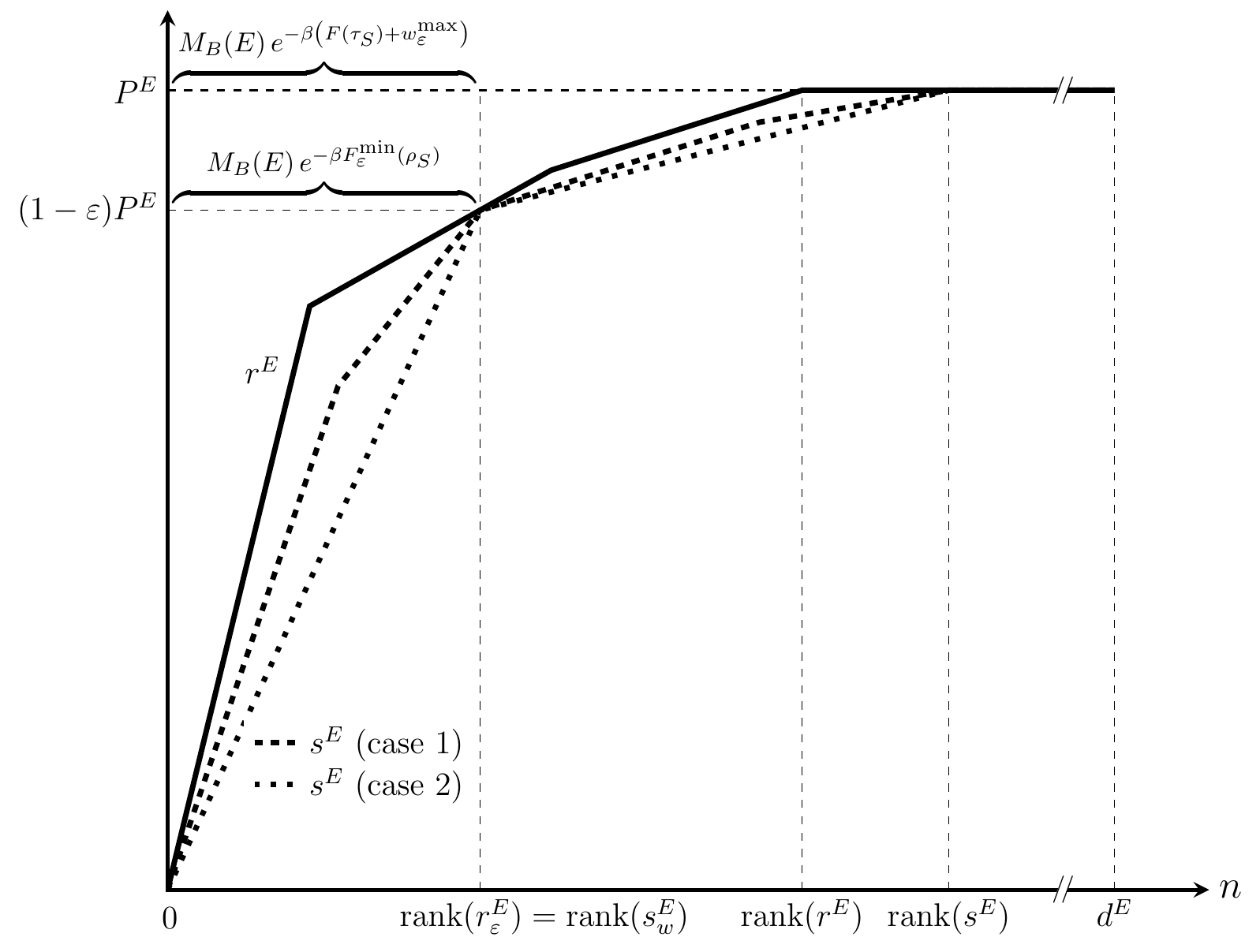}
\end{center}
\caption{Schematic Lorenz curves of $r^E$ and two possible cases of $s^E$ for imperfect maximum work extraction, and graphical interpretation of the single-shot free energy $F^\mathrm{min}_\varepsilon(\rho_S)$ and maximum extractable work $w^\mathrm{max}_\varepsilon$. Note that $s^E$ has full rank in the projective subspace $\Pi_W(0)+\Pi_W(w)$, $s^E_w$ has full rank in the projective subspace $\Pi_W(w)$, and $r^E_\varepsilon$ and $s^E_w$ have the same rank (or dimension). The horizontal axis is broken at $n\gtrsim\rk(s^E)$ for better illustration.}
\label{fig:fig2}
\end{figure}

The global unitary $V$ corresponds to maximum work extraction is not unique, as noted in Ref.~\cite{Gemmer:2015}. This is because any global unitary $V$ that transfers $r^E_\varepsilon$ entirely to $s^E_w$ such that $s^E_w$ has the same dimension as, and is majorized by, $r^E_\varepsilon$ will give rise to maximum work extraction, provided that $s^E_w$ has full rank in the projective subspace $\Pi_W(w)$. 
Hence, maximum work extraction can result in a multitude of final system states $\sigma_S$, and the decohered final system states $\mathcal{D}(\sigma_S)$ all have full rank.
A schematic of the Lorenz curves of $r^E$ and two possible cases of $s^E$ for imperfect maximum work extraction is shown in Fig.~\ref{fig:fig2}, where the final system state $\sigma_S$ in case 1 is a nonthermal state, while that in case 2 is the thermal state $\tau_S$. 
Since the final system state $\sigma_S$ is in general not diagonal in the system energy eigenbasis, if we want to apply a second work extraction to the state $\sigma_S$ (presumably with a ``fresh bath'' that is uncorrelated with the system), it needs to be decohered so as to remove the off-diagonal elements~\cite{Horodecki:2013}.
Therefore, because all of the decohered final system states after a maximum work extraction have full rank, it is \emph{not} possible to extract more work with certainty (i.e., $\varepsilon=0$) by transforming a nonthermal final system state to another state, be it thermal or otherwise. 

If the final system state after a maximum work extraction is a \emph{nonthermal} state $\sigma_S$, it seems that even more work may be extracted using a second, \emph{imperfect} work extraction. However, as we will argue below, this is not the case. 
Our arguments rest on the fundamental concepts of coherence work-locking, quantum typicality, and complete passivity, and strengthen the arguments given in Ref.~\cite{Gemmer:2015}.
Note that before the above analysis can be applied to find the extractable work (with failure probability $\varepsilon$), the final system state $\sigma_S$ needs to be decohered so as to remove the off-diagonal elements.
Nevertheless, the maximum extractable work of the state $\sigma_S$ is the same as that of the decohered state $\mathcal{D}(\sigma_S)$ because of coherence work-locking~\cite{Lostaglio:2015,Korzekwa:2016} (see also Refs.~\cite{Horodecki:2013,Skrzypczyk:2014}). 
The latter refers to the phenomenon that, under thermal operations (or strictly energy-conserving global unitaries) and without additional coherence resources, coherence of individual quantum systems contributes to the free energy but not extractable work. 
Hence, the maximum extractable work of the state $\sigma_S$ is determined by its diagonal part $\mathcal{D}(\sigma_S)$.
In the context of quantum typicality~\cite{Goldstein:2006,Reimann:2007,Popescu:2006}, it has been shown in Ref.~\cite{Gemmer:2015} that for an energy-conserving global unitary $V$ drawn at random, a typical final system state $\sigma_S$ is element-wise close to the thermal state $\tau_S$, if the bath is large enough but finite, with relative fluctuations on the order of $1/\sqrt{M_B(E)}\ll 1$ in the energy shell $E$.
Here, as in Ref.~\cite{Gemmer:2015}, typicality is understood in the same sense as that is usually used in the context of thermalization studies, i.e., a state is typical if it is overwhelmingly frequent with respect to Haar-distributed global unitaries~\cite{Goldstein:2006,Reimann:2007,Popescu:2006}.
We hasten to note that because work is not a state function, even if the typical final system state is element-wise close to the thermal state, the work extracted in each single run of the extraction process may differ significantly. 
While the typical final system state $\sigma_S$ is nonthermal and coherent, the decohered state $\mathcal{D}(\sigma_S)$ is close to the thermal state $\tau_S$, i.e., we have $\mathcal{D}(\sigma_S)\approx\tau_S$. 
For the second, imperfect work extraction, the initial global state is then given by $\eta\coloneqq\mathcal{D}(\sigma_S)\otimes\tau_B\otimes|0\rangle_W\langle 0|\approx\tau_S\otimes\tau_B\otimes|0\rangle_W\langle 0|$.
It is well known that for a given system, the thermal state is a passive state, and the only completely passive state, of the system~\cite{Pusz:1978,Lenard:1978}. 
A passive state is defined to be a state that does not allow for work extraction via reversible unitary processes, while a completely passive state $\rho$ is a state for which the global state $\rho^{\otimes N}$ of $N$ independent and identically distributed copies of the system is passive for all positive integers $N$.
Very recently, it has been shown~\cite{Sparaciari:2017} that an engine utilizing an ancillary system can extract a maximum amount of work from a single copy of any passive state, but not completely passive state, and in order to extract work from passive states close to (in trace norm) the set of completely passive states, an ancillary system of large dimension is needed. 
Hence, we conclude that in the absence of an ancillary system, a second, imperfect work extraction cannot extract any work from the final system state $\sigma_S$ of a previous maximum work extraction. This explains why there is no second law violation with the final system state of maximum work extraction being nonthermal.

\section{Minimum work cost of formation}\label{sec:workformation}

The bounds on the work cost of formation from a thermal system state can also be analyzed using our approach. Unlike the case of work extraction, we will first derive the minimum work cost for \emph{exact} formation of a nonthermal state $\sigma_S$ from a thermal state $\tau_S$ under thermal operations, and then use the result to derive the minimum work cost for \emph{approximate} formation, in a given distance measure, of states $\varepsilon$-close to the desired target state $\sigma_S$ for some fixed $\varepsilon$.
This is because of the difference in the nature of uncertainty associated with single-shot work extraction and state formation. The uncertainty in the former case is of probabilistic nature, i.e., work extraction with certainty will be impossible for system states that have full rank, and hence a failure probability of $\varepsilon$ is acceptable; however, the uncertainty in the latter case is of metrological nature, i.e., it will be practically impossible to exactly determine a target state with infinite accuracy, and hence an error of $\varepsilon$ is tolerable.   

We will consider thermal operations associated with the following unitary transformation 
\begin{equation}
\eta\coloneqq\tau_S\otimes\tau_B\otimes|w\rangle_W\langle w|\xrightarrow{~V~}\sigma_{S}\otimes\tau_B\otimes |0\rangle_W\langle 0|\eqqcolon\eta',
\end{equation}
where, again, the global unitaries $V$ commute with the total Hamiltonian $H$. The final state of the system and bath is taken to be a product state $\sigma_S\otimes\tau_B$ with the bath in the thermal state $\tau_B$. 
We stress that this choice of the final global state $\eta'$ ensures that the work cost is used solely to transform the system state from $\tau_S$ to $\sigma_S$, and thus is minimum \emph{per se}. The product state $\sigma_S\otimes\tau_B$ of the system and bath was considered in Ref.~\cite{Horodecki:2013}, and explicitly assumed at some point of the derivation in Ref.~\cite{Gemmer:2015}.

Following the same reasoning used in the case of work extraction, we find that the eigenvalues of the initial global state $\eta$ in the energy shell $E$ (i.e., the projective state $\eta^E$) are given by
\begin{equation}
r^E(E_S,g_S,f_B,E_W)=\frac{e^{-\beta(E-w)}}{Z_S Z_B}\delta_{E_W,w},\label{eq:reform}
\end{equation}
where $g_S=1,\ldots,M_S(E_S)$ and $f_B=1,\ldots,M_B(E-E_S-E_W)$ are the degeneracy indices of the system and bath at energies $E_S$ and $E_B=E-E_S-E_W$, respectively.  
Since the global unitary $V$ preserves probability, the eigenvalues of the final global state $\eta'$ in the energy shell $E$ (i.e., the projective state $\eta'^E$) are also given by $r^E(E_S,g_S,f_B,E_W)$.
The diagonal elements of the projective state $\eta'^E$ in the global energy eigenbasis are given by
\begin{equation}
s^E(E_S,g_S,f_B,E_W)=\langle E_S,g_S|\sigma_S|E_S,g_S\rangle\frac{e^{-\beta(E-E_S)}}{Z_B}\delta_{E_W,0},\label{eq:seform}
\end{equation}
where $g_S=1,\ldots,M_S(E_S)$ and $f_B=1,\ldots,M_B(E-E_S-E_W)$ are the same degeneracy indices as those for $r^E(E_S,g_S,f_B,E_W)$.

\begin{figure}[t]
\begin{center}
\includegraphics[width=4.25in]{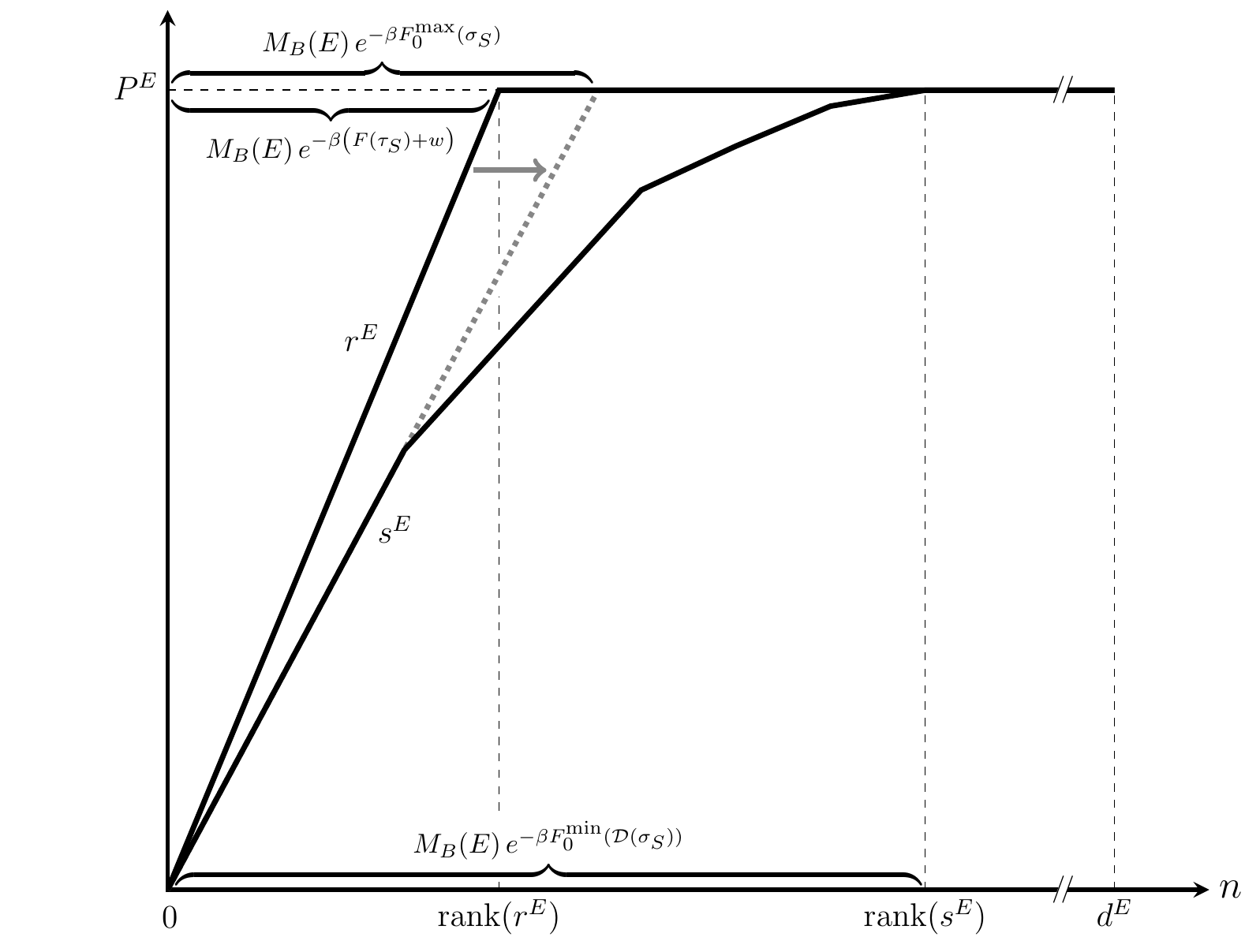}
\end{center}
\caption{Schematic Lorenz curves of $r^E$ and $s^E$ for exact state formation, and graphical interpretation of the work cost of formation $w$, single-shot free energies $F^\mathrm{max}_0(\sigma_S)$ and $F^\mathrm{min}_0(\mathcal{D}(\sigma_S))$. The gray arrow indicates the stretching of the Lorenz curve of $r^E$ along the $n$-axis such that its on-ramp is aligned with that of the Lorenz curve of $s^E$. The horizontal axis is broken at $n\gtrsim\rk(s^E)$ for better illustration.}
\label{fig:fig3}
\end{figure}

To apply the Schur theorem to the probability vectors $r^E$ and $s^E$, we need to find $r^{E\downarrow}$ and $s^{E\downarrow}$. 
Note that $r^E(E_S,g_S,f_B,E_W)$ are independent of $E_S$, $g_S$, and $f_B$, and nonzero only for $E_W=w$. This implies that the vector $r^E$ has rank equal to the dimension of the projective subspace $\Pi_W(w)$ in the energy shell $E$, and all of its nonzero components are equal. Hence, we have $r^{E\downarrow}_1=\cdots=r^{E\downarrow}_{\rk(r^E)}>r^{E\downarrow}_{\rk(r^E)+1}=\cdots=r^{E\downarrow}_{d^E}=0$. 
Because $s^E(E_S,g_S,f_B,E_W)$ in Eq.~\eqref{eq:seform} are formally the same as $r^E(E_S,g_S,f_B,E_W)$ in Eq.~\eqref{eq:reext}, we have $s^{E\downarrow}_1\ge\ldots\ge s^{E\downarrow}_{d^E}$, where $s^{E\downarrow}_1$ is the $s^E(E_S,g_S,f_B,0)$ with the largest $\langle E_S,g_S|\sigma_S|E_S,g_S\rangle e^{\beta E_S}$, $r^{E\downarrow}_2$ is that with the second largest $\langle E_S,g_S|\sigma_S|E_S,g_S\rangle e^{\beta E_S}$, and so on, until that with the smallest nonzero $\langle E_S,g_S|\sigma_S|E_S,g_S\rangle e^{\beta E_S}$ is reached, and then the remaining are all zeros. 
In other words, $s^E$ (or, equivalently, the vector of diagonal elements $\langle E_S,g_S|\sigma_S|E_S,g_S\rangle$) is $\beta$-ordered~\cite{Horodecki:2013}. A schematic of the Lorenz curves of $r^E$ and $s^E$ is shown in Fig.~\ref{fig:fig3}. 
It is easy to see that the Lorenz curve of $r^E$ has a simple form of an on-ramp (i.e., the first segment of the curve) and a tail (i.e., the right-most part where the curve is flat), and the horizontal width of the on-ramp (i.e., the rank of $r^E$) is determined by the dimension of the projective subspace $\Pi_W(w)$ in the energy shell $E$. 
In fact, this is a generic property of the Lorenz curve for any projective state $\eta^E$ with $\eta=\tau_S\otimes\tau_B\otimes|w\rangle_W\langle w|$, which we shall referred to as a \emph{thermal product state} of the system and bath at weight energy $w$, or simply thermal product state for short.

According to the Schur theorem, it follows that $r^E$ majorizes $s^E$. A lower bound on the work cost of formation $w$ can be immediately obtained from the implication that $\rk(r^E)\le\rk(s^E)$. Here, we have $\rk(r^E)=M_B(E)e^{-\beta(F(\tau_S)+w)}$ and $\rk(s^E)=M_B(E)e^{-\beta F^\mathrm{min}_0(\sigma_S)}$, where use has been made of Eq.~\eqref{eq:reform} and the analysis of Sec.~\ref{sec:workextraction}. However, the resulting lower bound is far from stringent. 
To find a more stringent lower bound, we first note the fact that $r^E$ corresponds to a thermal product state and $r^E$ majorizes $s^E$ implies $r^{E\downarrow}_1=\cdots=r^{E\downarrow}_{\rk(r^E)}\ge s^{E\downarrow}_1\ge\cdots\ge s^{E\downarrow}_{\rk(s^E)}>0$; hence, the on-ramp slope of the Lorenz curve of $r^E$ is greater than or equal to that of $s^E$. 
We can then implement a $\gamma$-fold ($\gamma\ge 1$) stretching of the Lorenz curve of $r^E$ along the $n$-axis (making its on-ramp slope decreased by a factor of $\gamma$), and still find that the Lorenz curve of $r^E$ is everywhere greater than or equal to that of $s^E$ (see Fig.~\ref{fig:fig3}). Thus, from Eqs.~\eqref{eq:reform} and \eqref{eq:seform}, we obtain
\begin{equation}
w\ge kT\log\bigl(\gamma\,\langle E_S^\ast,g_S^\ast|\sigma_S|E_S^\ast,g_S^\ast\rangle\,e^{\beta E_S^\ast}Z_S\bigr),
\end{equation}
where $|E_S^\ast,g_S^\ast\rangle$ is the system energy eigenstate corresponding to $s^{E\downarrow}_1$, i.e., the energy eigenstate with the largest $\beta$-ordered diagonal element of the target system state $\sigma_S$. 
The most stringent lower bound on $w$ is given by $\gamma=1$, corresponding to the on-ramps of the Lorenz curves of $r^E$ and $s^E$ are aligned in the first place, and reads
\begin{equation}
w\ge w^\mathrm{min}_0\coloneqq kT\log\bigl(\langle E_S^\ast,g_S^\ast|\sigma_S|E_S^\ast,g_S^\ast\rangle\,e^{\beta E_S^\ast}Z_S\bigr).\label{eq:wgewmin}
\end{equation}
To compare our result for the minimum work cost of formation $w^\mathrm{min}_0$ with those derived in the literature~\cite{Horodecki:2013,Gemmer:2015}, we note that the product under the logarithm on the right-hand side of Eq.~\eqref{eq:wgewmin} can be expressed as
\begin{equation}
\langle E_S^\ast,g_S^\ast|\sigma_S|E_S^\ast,g_S^\ast\rangle\,e^{\beta E_S^\ast}Z_S=\min\{\lambda:\sigma_S\le\lambda\,\tau_S\},\label{eq:minlambda}
\end{equation}
where use has been made of the relation $e^{-\beta E_S^\ast}/Z_S=\langle E_S^\ast,g_S^\ast|\tau_S|E_S^\ast,g_S^\ast\rangle$. 
It is convenient to define another single-shot free energy $F^\mathrm{max}_0(\sigma_S)$ by\footnote{In resource-theoretic terms, $F^\mathrm{max}_0(\sigma_S)$ is called the max-free energy of the state $\sigma_S$~\cite{Horodecki:2013}, a quantity related to the max-relative entropy of $\sigma_S$ to $\tau_S$~\cite{Gour:2015,Datta:2009}, and hence the notation. Its smoothed version is the $\varepsilon$-smooth max-free energy $F^\mathrm{max}_\varepsilon(\sigma_S)$ defined in Eq.~\eqref{eq:fmaxep}.}
\begin{equation}
F^\mathrm{max}_0(\sigma_S)\coloneqq kT\log\min\{\lambda:\sigma_S\le\lambda\,\tau_S\}+F(\tau_S),
\end{equation}
in terms of which the minimum work cost of formation $w^\mathrm{min}_0$ in Eq.~\eqref{eq:wgewmin} can be written as
\begin{equation}
w^\mathrm{min}_0=F^\mathrm{max}_0(\sigma_S)-F(\tau_S).\label{eq:wgewminfmax}
\end{equation} 
Again, as shown in Fig.~\ref{fig:fig3}, our approach allows for a graphical interpretation of the work cost of formation $w$ and single-shot free energy $F^\mathrm{max}_0(\sigma_S)$ in terms of the Lorenz curves of $r^E$ and $s^E$. An inspection of Fig.~\ref{fig:fig3} reveals that $w^\mathrm{min}_0$ is indeed the greatest lower bound of $w$.
Note that $w^\mathrm{min}_0$ is independent of the energy shell $E$ because the trivial dependence of $e^{-\beta E}$ on $E$ in $r^E$ and $s^E$ cancels out; hence, the minimum work cost of formation for exact formation previously derived in Refs.~\cite{Horodecki:2013,Gemmer:2015} is obtained. 
Moreover, the minimum work cost of formation $w^\mathrm{min}_0$ can be achieved if and only if, in any energy shell $E$, the largest diagonal elements of the final projective state $\eta'^E$ in the projective subspace $\Pi_W(0)$ is the same as the (degenerate) eigenvalue of the initial projective state $\eta^E$ in the projective subspace $\Pi_W(w^\mathrm{min}_0)$.

For approximation formation of states $\varepsilon$-close to the target state $\sigma_S$ in trace norm, the minimum work cost of formation $w^\mathrm{min}_\varepsilon$ can be found by minimizing $F^\mathrm{max}_0(\sigma'_S)$ with respect to all states $\sigma'_S$ in the $\varepsilon$-vicinity of the state $\sigma_S$ in trace norm. Thus, we obtain 
\begin{equation}
w\ge w^\mathrm{min}_\varepsilon\coloneqq F^\mathrm{max}_\varepsilon(\sigma_S)-F(\tau_S).\label{eq:wgewminepfmax}
\end{equation} 
Here, $F^\mathrm{max}_\varepsilon(\sigma_S)$ is the corresponding single-shot free energy defined by
\begin{equation}
F^\mathrm{max}_\varepsilon(\sigma_S)\coloneqq\inf_{\sigma'_S}F^\mathrm{max}_0(\sigma'_S),\label{eq:fmaxep}
\end{equation}
where the infimum is taken over all $\sigma'_S$ satisfying $\Vert\sigma'_S-\sigma_S\Vert_1\le\varepsilon$ with $\Vert\sigma\Vert_1\coloneqq\tr\sqrt{\sigma^\dagger\sigma}$ being the trace norm~\cite{Wilde:2017}. Clearly, the minimum work cost of formation for approximate formation previously derived in Refs.~\cite{Horodecki:2013,Gemmer:2015} is recovered.

Moreover, an inspection of Fig.~\ref{fig:fig3} reveals that $F^\mathrm{max}_0(\sigma_S)\ge F^\mathrm{min}_0(\mathcal{D}(\sigma_S))\ge F(\tau_S)$ and, in general, $F^\mathrm{max}_\varepsilon(\sigma_S)\ge F^\mathrm{min}_\varepsilon(\mathcal{D}(\sigma_S))\ge F(\tau_S)$ for approximate state formation and imperfect work extraction, provided that $\varepsilon$ is not large. 
Hence, we cannot extract more work from a nonthermal state (a resource state in the resource theory of athermality~\cite{Brandao:2013}) than that is required to create the state, a restatement of the second law reflecting the fundamental irreversibility in thermodynamic processes.

\section{General work extraction}\label{sec:generalworkextraction}

Our approach can be generalized straightforwardly to the case of work extraction in which transitions from the ground state to multiple energy eigenstates in a certain energy range of the work storage system are considered successful~\cite{Gemmer:2015}. 
Specifically, the amount of extracted work $\widetilde{w}_\varepsilon$ is now identified with the transition (with failure probability $\varepsilon$) of the work storage system from its ground state $|0\rangle_W$ to any state $|w\rangle_W$ in the energy range $\widetilde{w}\le w\le\widetilde{w}+\Delta$, where $\widetilde{w},\Delta>0$. 
Accordingly, the final state of the work storage system now takes the form
\begin{equation}
\sigma_W=\varepsilon|0\rangle_W\langle 0|+(1-\varepsilon)\sum_{w=\widetilde{w}}^{\widetilde{w}+\Delta}p(w)|w\rangle_W\langle w|,\label{eq:sigmaWgeneral}
\end{equation}
where $p(w)\ge 0$ is the work distribution~\cite{Popescu:2006} that, given a successful work extraction process, characterizes the conditional probabilities of the weight in the state $|w\rangle_W$, and satisfies $\sum_{w=\widetilde{w}}^{\widetilde{w}+\Delta}p(w)=1$. 
A related problem concerning general work extraction with bounded fluctuations in work that is akin to the final weight state $\sigma_W$ in Eq.~\eqref{eq:sigmaWgeneral}, but without the failure probability $\varepsilon$, has been recently studied in Ref.~\cite{Richens:2016}.  

The analysis presented in Sec.~\ref{sec:workextraction} requires only minimum modification: (i) $w$ is not a fixed parameter, but taking values in the range $\widetilde{w}\le w\le \widetilde{w}+\Delta$, and (ii) $s^E_w$ is replaced by $s^E_{\widetilde{w}}$, which is the full-rank subnormalized probability vector whose components are nonzero $s^E(E_S,g_S,f_B,w)$ in decreasing order, i.e., the nonzero probabilities of the global energy eigenstates in the energy shell $E$ after a successful work extraction process. 
Clearly, the components of $s^E_{\widetilde{w}}$ add up to $(1-\varepsilon)P^E$.
Again, the fact that $r^E$ majorizes $s^E$ implies
\begin{equation}
\rk(r^E_\varepsilon)\le\rk(s^E_{\widetilde{w}}),\label{eq:rkrelesewt}
\end{equation}
where $\rk(s^E_{\widetilde{w}})$ is upper bounded by the number of the diagonal elements $s^E(E_S,g_S,f_B,w)$ or, equivalently, the dimension of the projective subspace $\sum_{w=\widetilde{w}}^{\widetilde{w}+\Delta}\Pi_W(w)$. Hence, we have
\begin{align}
\rk(s^E_{\widetilde{w}})&\le\sum_{w=\widetilde{w}}^{\widetilde{w}+\Delta}\sum_{E_S,g_S}M_B(E-E_S-w)\nn\\
&=M_B(E)Z_S\sum_{w=\widetilde{w}}^{\widetilde{w}+\Delta}e^{-\beta w}.\label{eq:rkrsewt}
\end{align}
Upon combining Eqs.~\eqref{eq:rankreeps}, \eqref{eq:rkrelesewt}, and \eqref{eq:rkrsewt}, we obtain
\begin{equation}
\widetilde{w}\le\widetilde{w}^\mathrm{max}_\varepsilon\coloneqq w^\mathrm{max}_\varepsilon+kT\log\sum_{w=\widetilde{w}}^{\widetilde{w}+\Delta}e^{-\beta(w-\widetilde{w})},\label{eq:wtildemaxep}
\end{equation}
where the trivial dependence of $M_B(E)$ on $E$ cancels out. 
The maximum extractable work $\widetilde{w}^\mathrm{max}_\varepsilon$ does not depend on the work distributions $p(w)$; however, it in general depends on the width of the allowed energy range $\Delta$ and the energy level spacings of the work storage system. The independence of $\widetilde{w}^\mathrm{max}_\varepsilon$ on $p(w)$ is not unexpected because it is the marginal probability of $1-\varepsilon$ that determines $\widetilde{w}^\mathrm{max}_\varepsilon$. Moreover, the analysis presented in Sec.~\ref{sec:workextraction} also shows that $\widetilde{w}^\mathrm{max}_\varepsilon$ is indeed the least upper bound of $\widetilde{w}$.

While our result agrees with that derived in Ref.~\cite{Gemmer:2015}, it gives rise to a serious conceptual inconsistency overlooked in the literature. It is evident that $\widetilde{w}^\mathrm{max}_\varepsilon\ge w^\mathrm{max}_\varepsilon$, and $\widetilde{w}^\mathrm{max}_\varepsilon$ in general increases with the number of energy levels in the target energy range for some fixed $\Delta$. Hence, an unlimited amount of work may be extracted from the system state $\rho_S$ by simply increasing the number of energy levels in the target energy range of the weight, leading to a violation of the second law.

To resolve this inconsistency, we note that the additional contribution to $\widetilde{w}^\mathrm{max}_\varepsilon$ in Eq.~\eqref{eq:wtildemaxep} is \emph{not} an intrinsic property of the system state $\rho_S$, because it is independent of $\rho_S$. In fact, it arises from the global energy eigenstates in the projective subspace $\sum_{w=\widetilde{w}}^{\widetilde{w}+\Delta}\Pi_W(w)$ (and the energy shell $E$) that are available to the total system when the final weight state is allowed to populate energy levels in the range  $\widetilde{w}\le w\le \widetilde{w}+\Delta$.
Hence, the crucial question as to what is the physical meaning of this additional contribution needs to be answered. To gain some insight into this question, let us consider perfect maximum work extraction from a system state $\rho_S$ that has full rank. As we have discussed in Sec.~\ref{sec:workextraction}, no work can be extracted from $\rho_S$ with certainty ($\varepsilon=0$), i.e., $w^\mathrm{max}_0(\rho_S)=0$. 
However, using an analysis similar to that presented in Sec.~\ref{sec:maxwork}, or simply setting $\varepsilon=0$ in Eq.~\eqref{eq:wtildemaxep}, we find 
\begin{equation}
\widetilde{w}^\mathrm{max}_0(\rho_S)=kT\log\sum_{w=\widetilde{w}}^{\widetilde{w}+\Delta}e^{-\beta(w-\widetilde{w})},
\end{equation}
which is nonnegative and, even worse, independent of $\rho_S$. Thus, by repeating this perfect maximum work extraction process indefinitely for some fixed $\Delta>0$, an infinite amount of work can be extracted \emph{with certainty} from any full-rank system state $\rho_S$, even the thermal system state $\tau_S$. 
In fact, the same arguments apply to the thermal system state $\tau_S$. Evidently, this is in contradiction with the complete passivity of the thermal state~\cite{Pusz:1978,Lenard:1978} and the second law. 
The situation becomes even more problematic for the simple case of a harmonic oscillator-like work storage system with equidistant energy spacing $\delta$, which was also considered in Ref.~\cite{Gemmer:2015}. From Eq.~\eqref{eq:wtildemaxep}, we find
\begin{equation}
\widetilde{w}^\mathrm{max}_\varepsilon=w^\mathrm{max}_\varepsilon+kT\log\frac{1-e^{-\beta(\Delta+\delta)}}{1-e^{-\beta\delta}},\label{eq:maxepharmosc}
\end{equation}
which in the limit of large energy intervals and small energy spacings (i.e., $\Delta\gg kT\gg\delta$) reduces to
\begin{equation}
\widetilde{w}^\mathrm{max}_\varepsilon=w^\mathrm{max}_\varepsilon+kT\log\frac{kT}{\delta}.\label{eq:maxepharmosclim}
\end{equation}
Clearly, $\widetilde{w}^\mathrm{max}_\varepsilon$ is sensitive to the details of the energy levels and diverges in the limit $\delta\to 0$. 
This not only renders the energy transfer unpredictable, but also signals a violation of the second law. 
Moreover, as can be seen from Eqs.~\eqref{eq:maxepharmosc} and \eqref{eq:maxepharmosclim}, the additional contribution to $\widetilde{w}^\mathrm{max}_\varepsilon$ is independent of the initial state of the weight, provided that the energy levels of the weight are equally spaced in the entire range $0\le w<\infty$, a necessary condition for the requirement that the global unitaries $V$ commute with translations on a weight with discrete energy levels (see Sec.~\ref{sec:preliminaries}). 
However, we stress that the independence of the extracted work on the initial state of the weigh is a necessary, but not sufficient, condition to ensure that the weight cannot be used as an entropy sink. 
Indeed, it has been noted in Ref.~\cite{Gallego:2016} that to quantify work for transitions of the weight involving multiple energy eigenstates of the form (i.e., a \emph{successful} general work extraction in the current context)
\begin{equation}
|0\rangle_W\langle 0|\longrightarrow\sum_w p(w)|w\rangle_W\langle w|,
\end{equation}
one has to properly account for the fact that the weight might act as an entropy sink, because there is an associated increase in the entropy of the weight.
Therefore, based on the above analysis, we conclude that the additional contribution to $\widetilde{w}^\mathrm{max}_\varepsilon$ in Eq.~\eqref{eq:wtildemaxep} should \emph{not} be interpreted as work extracted from the system, and hence the maximum extractable work from the system state $\rho_S$ remains unchanged, i.e., $\widetilde{w}\le w^\mathrm{max}_\varepsilon(\rho_S)$ [see Eq.~\eqref{eq:wlewmax}]. 

The careful reader may have noticed that the additional contribution to $\widetilde{w}^\mathrm{max}_\varepsilon$ in Eq.~\eqref{eq:wtildemaxep} is somehow related to the ``free energy'' of the work storage system \emph{as if} the latter is in the thermal state $\tau_W$ at temperature $T$, but with its energy confined in the range $\widetilde{w}\le w\le \widetilde{w}+\Delta$. Since the initial weight state is the ground state $|0\rangle_W\langle 0|$, which is a pure state and has vanishing free energy, the change in the free energy of the weight at the end of a successful work extraction is then given by
\begin{align}
\Delta F_W&=-kT\log\sum_{w=\widetilde{w}}^{\widetilde{w}+\Delta}e^{-\beta w}\nn\\
&=\widetilde{w}-kT\log\sum_{w=\widetilde{w}}^{\widetilde{w}+\Delta}e^{-\beta(w-\widetilde{w})}.\label{eq:deltafw}
\end{align}
Intriguingly, the free energy difference of the weight $\Delta F_W$ is the ``transfer quantity'' defined in Ref.~\cite{Gemmer:2015} for a more general case of work extraction in which transitions between multiple energy levels of the work storage system are consider successful. 
Because the standard formalism of thermal operations requires only strict energy conservation of the total system, but does not forbid energy exchange between the bath and the work storage system, it is conceivable that allowing the final and/or initial states of the weight to occupy multiple energy levels in a certain finite energy range will inevitably cause a spreading of energy across the different levels in the energy range. 
This in turn leads to an additional change in the free energy of the weight. Hence, the free energy change of the weight $\Delta F_W$ given by Eq.~\eqref{eq:deltafw} contains an artifact caused by the heat bath--work storage system coupling that is implicitly assumed in the standard formalism of thermal operations.
If only transitions from the ground state to single excited states of the weight are consider successful (i.e., the case studied in Secs.~\ref{sec:workextraction} and \ref{sec:maxwork}) then the free energy change of the weight would be the same as the energy stored in the weight, even though the bath and work storage system are implicitly assumed to be coupled to each other.
Thus, to remove this artifact we impose the condition $\Delta F_W=\widetilde{w}$ (or, equivalently, the requirement that the free energy change of the weight is exactly the energy it stores despite its coupling to the heat bath), and obtain the \emph{identity}
\begin{equation}
\log\sum_{w=\widetilde{w}}^{\widetilde{w}+\Delta}e^{-\beta(w-\widetilde{w})}=0.
\end{equation}
The physical meaning of this identity is that if we count only the work stored in the weight, then the additional contribution to $\widetilde{w}^\mathrm{max}_\varepsilon$ in Eq.~\eqref{eq:wtildemaxep} has to be identically zero or, more precisely, \emph{completely excluded}. Clearly, an obvious way to satisfy the identity is to take the limit $\Delta\to 0$, i.e., the limit of infinite precision, as is done in Ref.~\cite{Aberg:2013} to formalize the idea of an energy extraction that is essentially free from fluctuations. Indeed, it is exactly the quest for predictable ``truly'' work-like energy extraction from a quantum system that motivated the study of single-shot work extraction~\cite{Aberg:2013}.
To put it another way, the energy gain of the weight associated with the additional contribution to $\widetilde{w}^\mathrm{max}_\varepsilon$ is \emph{not} work extracted from the system \emph{but} heat transferred from the heat bath to the weight, a consequence that the weight does not distinguish work from heat. Hence, the conceptual inconsistency is clarified and resolved.
Our conclusion is in agreement with the result of Ref.~\cite{Egloff:2015} that only predictable energy transfer (with failure probability $\varepsilon$) should count as work, anything beyond that as heat, and allowing population in higher levels of the weight should not change the predicted work, or in our terms, maximum extractable work. 
Interestingly, the result therein is obtained in a \emph{distinct} setup in which a traditional thermodynamic definition of work is used without explicitly including a weight.

Nevertheless, we are left with two open questions: (i) how the standard formalism of thermal operations can be modified to include a work storage system that truly distinguishes work from heat; and (ii) what is an appropriate notion of work that captures the realistic operation of quantum thermal machines whose work storage system may have a large density of states.
While the second question is a part of a long-standing issue in quantum thermodynamics~\cite{Frenzel:2014,Gallego:2016}, the first question is, to the best of our knowledge, raised for the first time in the present article. 

Before ending this section, motivated by imperfect work extraction and approximate state formation, we consider a realistic and physically interesting generalization to approximate, imperfect work extraction with a failure probability of $\varepsilon$ from states $\delta$-close to the desired initial system state $\rho_S$ for some fixed $\delta$ in trace norm. 
According to the analysis presented in Secs.~\ref{sec:workextraction} and \ref{sec:workformation}, the maximum extractable work $w^\mathrm{max}_{\varepsilon,\delta}$ can be found by maximizing $F^\mathrm{min}_\varepsilon(\rho'_S)$ with respect to all diagonal states $\rho'_S$ in the $\delta$-vicinity of the state $\rho_S$ in trace norm. Hence, we have
\begin{equation}
w\le w^\mathrm{max}_{\varepsilon,\delta}\coloneqq F^\mathrm{min}_{\varepsilon,\delta}(\rho_S)-F(\tau_S),\label{eq:wlewmaxdelep}
\end{equation}
where $F^\mathrm{min}_{\varepsilon,\delta}(\rho_S)$ is the generalized single-shot free energy defined by\footnote{Incidentally, $F^\mathrm{min}_{\varepsilon,\delta}(\rho_S)$ in resource-theoretic terms would be the \emph{doubly smoothed} version of the min-relative entropy of $\rho_S$ to $\tau_S$.}
\begin{equation}
F^\mathrm{min}_{\varepsilon,\delta}(\rho_S)\coloneqq\sup_{\rho'_S}F^\mathrm{min}_\varepsilon(\rho'_S)\label{eq:fmaxdelep}
\end{equation}
with the supremum taken over all states $\rho'_S$ diagonal in the system energy eigenbasis and satisfying $\Vert\rho'_S-\rho_S\Vert_1\le\delta$. 

\section{Conclusions and Outlook}\label{sec:conclusions}

In this work, using a physically intuitive and mathematically simple approach, we have studied the problem of work extraction from a system in contact with a heat bath to a work storage system, and the reverse problem of state formation from a thermal system state in single-shot quantum thermodynamics.
Our approach uses only elementary majorization theory and matrix analysis, and builds a bridge between two previous methods based respectively on the concept of thermomajorization~\cite{Horodecki:2013} and a comparison of subspace dimensions~\cite{Gemmer:2015}.
The maximum extractable work and minimum work cost of formation are derived, and a graphical interpretation of the maximum extractable work, minimum work cost of formation, and corresponding single-shot free energies is presented.
Our approach also facilitates a detailed discussion of the necessary and sufficient conditions for both perfect and imperfect maximum work extraction. 
A seeming violation of the second law concerning the final system state being nonthermal after a maximum work extraction is resolved using arguments based on the fundamental concepts of coherence work-locking~\cite{Horodecki:2013,Skrzypczyk:2014,Lostaglio:2015,Korzekwa:2016}, quantum typicality~\cite{Goldstein:2006,Reimann:2007,Popescu:2006}, and complete passivity~\cite{Pusz:1978,Lenard:1978,Sparaciari:2017}. The arguments also strengthen those in the literature~\cite{Gemmer:2015} based solely on quantum typicality.
Indeed, it is because of coherence work-locking, i.e., the phenomenon that under thermal operations and without additional coherence resources, quantum coherence contributes to the free energy but not extractable work, that the initial system states can be assumed to be diagonal in the system eigenbasis.

This approach can be generalized to the case of general work extraction in which transitions from the ground state to multiple energy eigenstates in a certain energy range of the work storage system are considered successful~\cite{Gemmer:2015}. 
The resulting maximum extractable work contains an additional contribution that is independent of the initial system state and increases without bound as the density of states of the work storage system increases, thus leading to a violation of the second law and a conceptual inconsistency. 
A detailed analysis shows that the additional contribution should be interpreted \emph{not} as work extracted from the system \emph{but} as heat transferred from the heat bath to the weight, and hence the maximum extractable work from a system state remains unchanged in the general case. 
This conclusion agrees with a previous result obtained in a distinct setup in which a traditional thermodynamic definition of work is used and a weight is not explicitly included~\cite{Egloff:2015}.
The conceptual inconsistency is resolved by recognizing that the additional contribution is an artifact of the standard formalism of thermal operations, in which a coupling between the heat bath and the work storage system is implicitly assumed. 
To put it another way, if the formalism of thermal operations is generalized straightforwardly to include a work storage system, as is usually done in the literature, then under global unitaries that strictly conserve the total energy, the heat bath, system, and work storage system are inevitably all coupled together.
Thus, our result calls into question the naive notion that work in quantum thermodynamics can be reliably defined as the change in the energy of a weight.

There are several open questions to be answered. First, considering the wide use of a suspended weight as the work storage system in the thermal operation formalism~\cite{Horodecki:2013,Skrzypczyk:2014,Gemmer:2015,Richens:2016}, an improved work storage system that can distinguish work from heat  is of particular importance. 
The underlying rationale for thermal operations is to describe fundamental limitations to thermal transitions in their full generality; hence, restrictions on the class of allowed operations are kept minimum and any Hamiltonians of, as well as any couplings between, the heat bath and the system are allowed. However, a caveat to this freedom is that no physical principles shall be violated. 
Clearly, this is not the case when a work storage system (essentially a suspended weight) that does not distinguish work from heat is explicitly included. As we have shown, the implicitly assumed coupling between the bath and the work storage system could lead to a violation of the second law if special care is not taken.
While it is not clear at the moment what kinds of work storage systems in quantum thermodynamics are able to distinguish work from heat, the idea of an ``information fuel tape''~\cite{Bennett:1982,Feynman:1996} or ``information battery''~\cite{Faist:2015} seems very interesting. 
Our results strongly suggest that this is an avenue worth pursuing, and the related study will certainly shed light on the problem of an appropriate notion of work in quantum thermodynamics.

Second, since the single-shot scenario is in general associated with a worst case scenario, chances are that single-shot work extraction would not be maximum. 
If a work extraction (with failure probability $\varepsilon$) applied to an initial system state $\rho_S$ is \emph{not} maximum, then a dephasing operation and second work extraction (with the same failure probability $\varepsilon$) can be applied to the final system state $\sigma_S$ to extract more work from the system. 
This process can be repeated indefinitely until either a maximum work extraction is achieved, or the eventual final system state is close to the thermal system state $\tau_S$. 
From the analysis presented above, it is conceivable that the maximum amount of extracted work in the combined process is given by $w^\mathrm{max}_\varepsilon(\rho_S)$, in accordance with the second law. Nevertheless, this would take us away from the single-shot regime to an intermediate regime somewhere in between the single-shot and many-runs regimes. 
The latter refers to the scenario in which a large number of runs are applied to single systems in the \emph{same} initial state~\cite{Korzekwa:2016}, which clearly is not the case here. It would be interesting to study work extraction in this intermediate regime.

Last but not least, in the present approach to work extraction the initial system states are assumed to be diagonal in the system energy eigenbasis; however, a similar approach that can be applied directly to initial system states with coherence in the system energy eigenbasis would be highly desirable. 
In particular, the generalized approach will provide a clear manifestation of coherence work-locking and how work is ``unlocked'' in the presence of additional coherence resources. 
Since the final system states of state formation are in general states with coherence in the system energy eigenbasis, the global unitary transformation $\eta\coloneqq\rho_S\otimes\tau_B\otimes{|0\rangle_W\langle 0|}\xrightarrow{\,V\,}\sigma_S\otimes\tau_B\otimes{|w\rangle_W\langle w|}\eqqcolon\eta'$ for work extraction from quantum states $\rho_S$ with coherence can be thought of as an inverse state formation followed by a forward one, i.e., $\eta\xrightarrow{\,V_1\,}\overline{\eta}\xrightarrow{\,V_2\,}\eta'$, where $\overline{\eta}\coloneqq\tau_S\otimes\tau_B\otimes{|w+w'\rangle_W\langle w+w'|}$ with $w,w'>0$, and $V_1$, $V_2$, and $V=V_2V_1$ are global unitaries. Here, for simplicity, we have assumed that the final state of the system and bath after the work extraction is a product state $\sigma_S\otimes\tau_B$. Hence, the global states $\eta$, $\eta'$, and $\overline{\eta}$ all have the same eigenvalues, which are given by the diagonal elements of $\overline{\eta}$. Investigation along this line is an interesting future avenue.
We believe that this article leads to a simple understanding of the underlying physics and provides a new insight into these open questions.

\acknowledgments

I would like to thank G.\ Auletta and C.N.\ Leung for a careful reading of the manuscript and useful comments, and M.P.\ M\"uller for sharing freely online his handwritten lecture notes on single-shot quantum thermodynamics, through which this fascinating field was first introduced to me. 
I would also like to thank the anonymous referees for useful comments and suggestions.

\end{document}